\documentclass[conference]{IEEEtran}
\IEEEoverridecommandlockouts

\usepackage{cite}
\usepackage{amsmath,amssymb,amsfonts}
\usepackage{algorithmic}
\usepackage{graphicx}
\usepackage{textcomp}
\usepackage{xcolor}

\usepackage{mwe}
\usepackage{url}
\usepackage{multirow}
\usepackage{enumitem}
\usepackage{util}
\usepackage{booktabs}
\usepackage{subfigure}

\def\BibTeX{{\rm B\kern-.05em{\sc i\kern-.025em b}\kern-.08em
    T\kern-.1667em\lower.7ex\hbox{E}\kern-.125emX}}

\begin{document}

\title{MambaITD: An Efficient Cross-Modal Mamba Network for Insider Threat Detection\\
}

\author{\IEEEauthorblockN{Kaichuan Kong}
\IEEEauthorblockA{\textit{College of Cyber Security} \\
\textit{Jinan University}\
Guangzhou, China \\
willkkc@stu2021.jnu.edu.cn}
\and
\IEEEauthorblockN{Dongjie Liu}
\IEEEauthorblockA{\textit{College of Cyber Security} \\
\textit{Jinan University}\\
Guangzhou, China \\
djliu@jnu.edu.cn}
\and
\IEEEauthorblockN{Xiaobo Jin}
\IEEEauthorblockA{\textit{Department of Electrical and Electronic Engineering} \\
\textit{Xi’an Jiaotong-Liverpool University}\\
Suzhou, China \\
xiaobo.jin@xjtlu.edu.cn}
\and
\IEEEauthorblockN{Zhiying Li}
\IEEEauthorblockA{\textit{College of Cyber Security} \\
\textit{Jinan University}\\
Guangzhou, China \\
tzezd2019@stu2020.jnu.edu.cn}
\and
\IEEEauthorblockN{Guanggang Geng}
\IEEEauthorblockA{\textit{College of Cyber Security} \\
\textit{Jinan University}\\
Guangzhou, China \\
gggeng@jnu.edu.cn}
\and
\IEEEauthorblockN{Jian Weng}
\IEEEauthorblockA{\textit{College of Cyber Security} \\
\textit{Jinan University}\\
Guangzhou, China \\
cryptjweng@gmail.com}
}

\maketitle

\begin{abstract}
Enterprises are facing increasing risks of insider threats, while existing detection methods are unable to effectively address these challenges due to reasons such as insufficient temporal dynamic feature modeling, computational efficiency and real-time bottlenecks and cross-modal information island problem. This paper proposes a new insider threat detection framework MambaITD based on the Mamba state space model and cross-modal adaptive fusion. First, the multi-source log preprocessing module aligns heterogeneous data through behavioral sequence encoding, interval smoothing, and statistical feature extraction. Second, the Mamba encoder models long-range dependencies in behavioral and interval sequences, and combines the sequence and statistical information dynamically in combination with the gated feature fusion mechanism. Finally, we propose an adaptive threshold optimization method based on maximizing inter-class variance, which dynamically adjusts the decision threshold by analyzing the probability distribution, effectively identifies anomalies, and alleviates class imbalance and concept drift. Compared with traditional methods, MambaITD shows significant advantages in modeling efficiency and feature fusion capabilities, outperforming Transformer-based methods, and provides a more effective solution for insider threat detection.
\end{abstract}

\begin{IEEEkeywords}
Mamba, Insider threat detection, Cross-modal fusion, Behavioral interval analysis, Adaptive threshold optimization
\end{IEEEkeywords}

\section{Introduction}
In recent years, the complexity and concealment of internal threats in enterprises have increased significantly, posing a severe challenge to organizational security and business continuity~\cite{gartner2023cyber}. According to the ``2024 Insider Threat Report" released by Cybersecurity Insiders~\cite{gurucul2024insider}, 83\% of enterprises have suffered at least one internal attack in the past year, while 52\% of enterprises admit that they lack effective defense tools. This contradiction shows that existing detection technologies are difficult to cope with new threat patterns and urgently need to break through the limitations of traditional methods. The shortcomings of current research are mainly reflected in the following three aspects:

\tf{Insufficient temporal dynamic feature modeling}: Mainstream methods (such as sequence modeling based on Transformer or LSTM~\cite{ergen2020unsupervised,pal2023temporal,villarreal2021hunting}) focus on the sequential nature of user behavior, but ignore the semantic value of the time interval between behaviors. For example, the time interval between intensive abnormal logins or low-frequency operations may imply attack intent, but the existing models do not explicitly model such features, resulting in limited threat perception sensitivity.

\tf{Computational efficiency and real-time bottlenecks}: Although methods based on pre-trained models such as BERT~\cite{guo2022logbert,huang2021itdbert} can capture behavioral semantics, their quadratic self-attention mechanism brings high computational overhead, making it difficult to meet the real-time processing requirements of large-scale log streams, limiting the feasibility of deployment in industrial scenarios.

\tf{Cross-modal information island problem}: Existing systems lack the ability to collaboratively analyze heterogeneous data sources (such as network traffic, operation logs, and resource status~\cite{inayat2024insider}), resulting in fragmented threat clues, inability to build a global risk profile, and weakening the situational awareness capability of the detection system.

To address these challenges, we present MambaITD, a novel framework that integrates Mamba-based state space models with cross-modal adaptive fusion for real-time insider threat detection: 1) Based on the Mamba network modeling of user behavior sequences, the time interval information between behaviors is explicitly encoded for the first time in internal threat detection, and the long-range dependency modeling capability of SSM is used to capture the temporal correlation of threat behaviors (such as the temporal clustering of intensive abnormal operations), overcoming the limitation of traditional methods that only focus on the order of events; 2) Mamba is used to replace traditional Transformer/BERT, and the linear complexity characteristics of the state space model are used to reduce the computational overhead (compared with the square complexity of Transformer), support real-time detection under large-scale data, and solve the bottleneck problems of high deployment cost and large response delay of existing deep learning models; 3) A gated feature fusion (GFF) mechanism is designed to dynamically aggregate behavior sequences, time intervals and multi-source statistical features based on statistical priors to break the ``information island"; an adaptive threshold optimization algorithm is further proposed to dynamically adjust the detection threshold through probability distribution analysis and inter-class variance maximization to improve the generalization ability in complex scenarios.

The main contributions of this work are as follows:
\begin{itemize}

\item \tf{Multi-source heterogeneous spatiotemporal feature reconstruction}: the first log fusion paradigm for behavioral interval perception, which aligns the scattered original logs (network traffic, system audit, etc.) in time and normalizes the granularity, extracts three-dimensional structured features (behavior sequence, interval sequence, statistical indicators), and breaks through the limitations of traditional single-dimensional event flow modeling;

\item \tf{Dual-channel temporal semantic embedding}: Through differentiated embedding strategies, behavioral semantics (such as operation type, object sensitivity) and temporal dynamics (such as operation interval density, abnormal time period distribution) are encoded respectively, and behavior-time coupling representation is constructed in low-dimensional space;

\item \tf{Lightweight state space modeling and dynamic fusion}: The state transition equation of the Mamba network is used to model long-range behavioral dependencies, and the threat aggregation pattern in the interval sequence (such as high-frequency data access late at night) is captured synchronously, and a statistical prior-driven gating fusion mechanism is designed to dynamically balance the contribution weights of behavioral semantics and temporal dynamics;

\item \tf{Probability-driven adaptive decision engine}: Based on the probability distribution characteristics of threat behavior, the detection threshold is dynamically optimized through the principle of maximizing inter-class separability, realizing the evolution of detection strategies from ``static rules" to ``environmental perception", effectively coping with the concept drift problem in enterprise operations.

\end{itemize}

The remainder of this paper is organized as follows. Section~\ref{sec:related} reviews related work on insider threat detection and discusses the relevance of the Mamba architecture. Section~\ref{sec:prelim} introduces key preliminaries, including exponential weighted moving averages, state space models, and Otsu’s thresholding method. Section~\ref{sec:framework} presents our proposed framework, covering data preprocessing, feature embedding, Mamba-based encoding, and the adaptive threshold optimization module. Section~\ref{sec:setup} outlines the experimental setup, including datasets, baselines, and evaluation metrics. Section~\ref{sec:results} reports experimental results with comparisons, ablation studies, and parameter analyses. Finally, Section~\ref{sec:conclusion} concludes the paper.

\section{Related Work}
\label{sec:related}

\subsection{Insider Threat Detection}
Early research on insider threat detection (ITD) primarily relied on manually defined statistical features (e.g., user login/logout frequency, email interaction volume) to construct feature sets, combined with traditional classifiers such as Hidden Markov Models (HMMs)~\cite{rashid2016new} and Isolation Forests~\cite{lv2018hybrid} for anomaly detection. Although these methods achieved moderate success in user-level anomaly screening, they exhibited significant limitations: the inability to model the temporal dynamics of user behavior hindered their capability to detect complex threats~\cite{gong2024graph,mahboubi2024evolving}.

To capture the complex dependencies between users and activities, researchers introduced graph neural networks (GNNs)~\cite{cai2024lan,li2023high}, modeling organizational interactions as communication or heterogeneous graphs. For instance, edge-centric anomaly scoring methods demonstrated preliminary effectiveness in heterogeneous graphs. However, these methods face two major challenges~\cite{gong2024graph}: first, the computational overhead increases quadratically with graph size, and second, heuristic graph construction methods may introduce human biases, limiting their applicability in real-world scenarios. Consequently, despite being a recent advancement, the high resource consumption of graph-based methods has shifted the research focus back to sequence modeling techniques.

Current research frontiers focus on sequence modeling techniques, which significantly enhance detection efficiency and accuracy by automatically learning the spatiotemporal evolution patterns of user behavior. This progression began with the integration of temporal point processes with LSTM~\cite{pal2023temporal,villarreal2021hunting} to identify anomalous segments in fine-grained behavior sequences. The Transformer model, with its powerful self-attention mechanism and parallel processing capabilities, has gradually been introduced into the field of ITD. For example, the use of Transformer encoders allows for the learning of temporal dependencies and patterns in user behavior sequences~\cite{xiao2024unveiling}. Additionally, pre-trained BERT models~\cite{huang2021itdbert} are employed to extract cross-session behavioral semantics, combined with bidirectional LSTMs to capture long-term dependencies.
 
Despite their strengths in capturing the temporal behavior of users, sequence modeling techniques still face the following shortcomings:
\textbf{Single-Perspective Modeling:} Most existing methods mainly focus on user behavior time series, overlooking the synergistic effects of temporal intervals between behaviors and statistical features, resulting in a narrow modeling perspective.
\textbf{Insufficient Feature Fusion:} Existing methods typically analyze behavioral sequences, temporal intervals, and statistical features independently before aggregating results. They fail to achieve deep integration of multi-scale features at the modeling level, which restricts further improvements in detection performance.

\subsection{Mamba for Detection}
Mamba is a novel sequence modeling architecture based on Structured State Space Models (SSM)~\cite{gu2023mamba}, specifically designed to address the limitations of traditional methods in anomaly detection~\cite{he2024mambaad,lei2024mambaad}. Compared to Transformer-based models, Mamba achieves linear complexity, enabling it to handle long sequences more efficiently—making it particularly suitable for insider detection tasks that require real-time responses~\cite{qu2024survey}. Its key features include linear scalability for processing large datasets and a mechanism for selective state updates, which allows it to dynamically focus on critical segments related to insider threats.

Research has shown that Mamba has broad applicability in the field of computer vision~\cite{li2024videomamba} and anomaly detection~\cite{pei2025enhancing}, especially in scenarios that require session feature modeling and low computational costs~\cite{liu2024mddmamba}. These characteristics position Mamba as an ideal choice for insider threat detection, enhancing its ability to detect local anomalies effectively.

\section{Preliminaries}
\label{sec:prelim}
\subsection{Exponential Weighted Moving Average}
The Exponential Weighted Moving Average (EWMA)~\cite{shi2023lstm} is a smoothing technique that assigns exponentially decreasing weights to past observations, making it particularly useful for capturing recent trends while mitigating noise. Given a time series \( \{z_1, z_2, ..., z_T\} \), the EWMA at time \( t \) is computed recursively as:  

\begin{equation}
q_t = \alpha z_t + (1 - \alpha) q_{t-1}, \quad 0 < \alpha \leq 1,
\end{equation}
where \( q_t \) represents the smoothed value at time \( t \), \( z_t \) is the raw observation, and \( \alpha \) is the smoothing factor that controls the decay rate of past values. A higher \( \alpha \) gives more weight to recent observations, making the model more responsive to changes, while a lower \( \alpha \) results in a smoother but slower-reacting trend.

\subsection{State Space Model}
The State Space Sequence Models (SSM)~\cite{gu2023mamba} designed to map a one-dimensional input sequence \( x(t) \in \mathbb{R} \) to an output sequence \( y(t) \in \mathbb{R} \). Its dynamics are governed by the following linear ordinary differential equations:  

\begin{equation}
h(t) = \mathbf{A} h(t) + \mathbf{B} x(t), \quad
\end{equation}

\begin{equation}
y(t) = \mathbf{C} h(t), \quad 
\end{equation}
where \( \mathbf{A} \in \mathbb{R}^{N \times N} \) and \( \mathbf{B}, \mathbf{C} \in \mathbb{R}^{N} \) are state matrices, and \( h(t) \in \mathbb{R}^{N} \) represents the hidden latent state.  

Structured state space models (SSM) have recently gained attention as a powerful class of sequence modeling architectures, demonstrating remarkable performance across various tasks. 

\subsection{Otsu's Thresholding}
\label{sec:otsu}

Otsu's method~\cite{goh2018performance} is a classical non-parametric algorithm originally designed for gray-level image segmentation. It aims to find an optimal threshold that separates a histogram into two classes by maximizing their inter-class variance. Given a normalized histogram $p(k)$ over $L$ discrete levels, the threshold $t^*$ is selected by:

\begin{equation}
t^* = \arg\max_t \left[ \omega_0(t)\omega_1(t)(\mu_0(t) - \mu_1(t))^2 \right],
\end{equation}

where $\omega_0(t)$ and $\omega_1(t)$ are the cumulative probabilities of the two classes, and $\mu_0(t)$ and $\mu_1(t)$ are their respective means. This strategy ensures that the separation between the two classes is statistically significant.

\section{Proposed Framework}
\label{sec:framework}

\begin{figure*}[!t]
\centering
\includegraphics[width=1.0\textwidth]{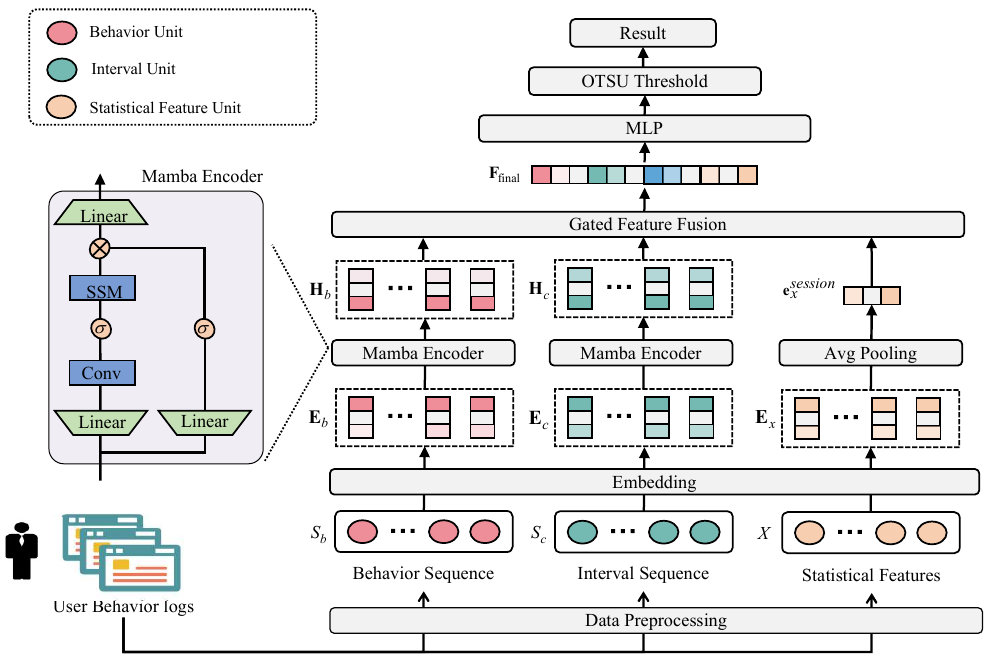}
\caption{The overview of MambaITD architecture. It uses the Mamba encoder to capture complex patterns in both behavioral and temporal features, which are then integrated through Gated Feature Fusion (GFF). An Otsu-based adaptive thresholding mechanism is applied for anomaly detection, classifying behaviors as either normal or anomalous.} 
\label{fig:architecture}
\end{figure*}

\subsection{Data Preprocessing}
The preprocessing module integrates multi-source heterogeneous logs and aligns temporal granularity across different sources. Since numerical scales and semantics vary across log sources, raw data cannot be directly used for subsequent analysis. Thus, this module transforms logs into a structured format to effectively model user behavior dynamics within a session. The feature generation process includes \textbf{behavior sequences}, \textbf{interval sequences}, and \textbf{statistical features}, which are essential for capturing user activity patterns.

\subsubsection{Behavior Sequence}
To model user behavior dynamics within a session, we encode each action using a Mapping ID, which integrates three key factors: Behavior, Device, and Time Segments~\cite{xiao2024unveiling}. The full encoding strategy and its detailed computation are presented in Appendix~\ref{appendix:behavior_encoding}.

For each session, the final Behavior Sequence is represented as:
\begin{equation}
S_b = [b_1, b_2, \ldots, b_T],
\end{equation}
where each \( b_i \) uniquely encodes the semantic and contextual aspects of user activity.

\subsubsection{Interval Sequence}
To capture temporal dynamics within a session, we construct an Interval Sequence that represents the time intervals between consecutive actions. To align the length of the Interval Sequence with the Behavior Sequence, we introduce a virtual initial interval \( c_0 = 0 \), representing the immediate occurrence of the first action at the session start time \( t_0 \). The raw time intervals \( \Delta t_i = t_i - t_{i-1} \) for \( i = 1, \ldots, T \) are smoothed using Exponential Weighted Moving Average (EWMA):
\begin{equation}
c_i = \alpha \Delta t_i + (1 - \alpha) c_{i-1}, \quad i = 1, \ldots, T,
\end{equation}
where \( \alpha = 0.2 \) controls the decay rate. The final Interval Sequence is given by:

\begin{equation}
S_c  = [c_1, c_2, \ldots, c_T].
\end{equation}

By truncating \( c_0 \), we obtain \( S_c \) with length \( T \), ensuring alignment with the Behavior Sequence. This approach captures both short-term and long-term temporal dependencies within a session.

\subsubsection{Statistical Features}
User behavior can be further characterized through statistical features across three dimensions: Behavior, Device, and Time Segments. These features capture the frequency and execution time characteristics of actions within these dimensions. Within a session, they provide a high-level summary of user activities, complementing the sequential and temporal information captured by the behavior sequence \( S_b \) and the interval sequence \( S_c \). The specific feature details are presented in Appendix~\ref{appendix:statistical_features}.

After standardization, these features form a numerical vector:
\begin{equation}
X = [x_1, x_2, ..., x_N],
\end{equation}  
where \( x_i \) represents the \( i \)-th feature value, reflecting the user's behavior within the corresponding dimension, and \( N \) is the total number of statistical features.

\subsection{Feature Embedding and Representation}
To effectively model user behavior, we construct three types of input data: the behavior sequence $ S_b $, the interval sequence $ S_c $, and the statistical feature vector $ X $. These heterogeneous features are projected into a unified latent space to facilitate joint representation learning, enabling the framework to capture multi-dimensional information for downstream tasks.

The embedding process is defined as follows:
\begin{align}
\mathbf{E}_b &= \text{Embed}(S_b), \\
\mathbf{E}_c &= \text{FC}(\log(1 + S_c)), \\
\mathbf{E}_x &= \text{BN}(\text{FC}(X)).
\end{align}

Here, $ \text{Embed}(\cdot) $ denotes a trainable embedding lookup table that maps each behavior ID $ b_i $ to a dense vector representation, resulting in $ \mathbf{E}_b \in \mathbb{R}^{T \times d_{\text{model}}} $. The function $ \text{FC}(\cdot) $ represents a fully connected layer that projects log-transformed interval values into the embedding space, producing $ \mathbf{E}_c \in \mathbb{R}^{T \times d_{\text{model}}} $. This transformation captures temporal dependencies while mitigating scale variations. For the statistical feature vector $ X $, which has $ N $ dimensions corresponding to the number of statistical features, $ \text{BN}(\cdot) $ applies batch normalization before mapping the features into a $ d_{\text{model}} $-dimensional space, resulting in $ \mathbf{E}_x \in \mathbb{R}^{N \times d_{\text{model}}} $. This ensures stability and consistency in feature distribution.

\subsection{Mamba Encoder and Feature Fusion}
\subsubsection{Mamba Block Encoder}
In the initial phase of Cross-Model Learning, the behavioral feature embeddings $ \mathbf{E_b} $ and temporal interval embeddings $ \mathbf{E_c} $ are fed into the Mamba framework to learn the hierarchical pattern information of the sequences. Mamba, a state-space model, excels at capturing long-term dependencies and complex patterns in sequential data. After processing through Mamba, we obtain two types of latent state representations:
\begin{align}
\mathbf{H}_b = \tm{Mamba}(\mathbf{E}_b) = [h_{b_1}; h_{b_2}; \dots; h_{b_T}] \in \mathbb{R}^{T \times d_{\text{model}}}, \\
\mathbf{H}_c = \tm{Mamba}(\mathbf{E}_c) = [h_{c_1}; h_{c_2}; \dots; h_{c_T}] \in \mathbb{R}^{T \times d_{\text{model}}}.
\end{align}

In this framework, each $ h_{b_i} $ and $ h_{c_i} $ represents a high-dimensional embedding derived from the long-term information modeling facilitated by Mamba. 

\subsubsection{Cross-Model Gated Feature Fusion}

\begin{figure}[!t]
\includegraphics[width=\linewidth]{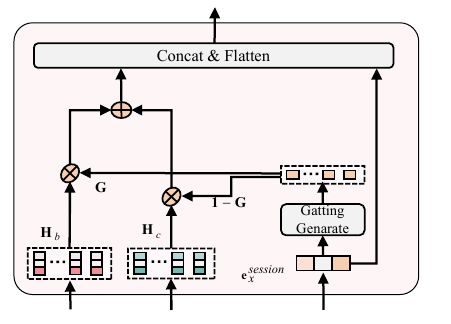}
\caption{Gated Feature Fusion mechanism. which integrates behavioral and temporal features (\( \mathbf{H}_b \) and \( \mathbf{H}_c \)) using a gating vector \( \mathbf{G} \). The vector is computed from session statistics (\( \mathbf{E}_x \)) and regulates the fusion of the two feature types. } 
\label{fig: Gatting}
\end{figure}

To effectively integrate behavioral latent states $ \mathbf{H}_b $ and temporal interval latent states $ \mathbf{H}_c $, we propose a Gated Feature Fusion (GFF) mechanism. This module dynamically adjusts the contributions of these features based on statistical priors, ensuring a balanced and informative fusion process.  

We first extract session statistical information from $ \mathbf{E}_x $ using average pooling:  
\begin{equation}
\mathbf{e}_x^{\text{session}} = \frac{1}{N} \sum_{i=1}^N \mathbf{E}_x[i,:].
\end{equation}

This vector captures overarching statistical patterns, such as mean frequency and interval variance, providing essential prior knowledge for feature fusion.  

\begin{equation}
\mathbf{G} = \sigma\left( \mathbf{W}_g \mathbf{e}_x^{\text{session}} + \mathbf{b}_g \right),
\end{equation}
where $ \sigma $ is the Sigmoid activation function, and $ \mathbf{W}_g $ and $ \mathbf{b}_g $ are trainable parameters. The gating vector $ \mathbf{G} $ assigns weights to each feature dimension, dynamically balancing behavioral and temporal interval information.  

The fused representation $ \mathbf{F}_{\text{fusion}} $ is then computed as:  
\begin{equation}
\mathbf{F}_{\text{fusion}}[t,j] = g_j h_{b_t,j} + (1 - g_j) h_{c_t,j},
\end{equation}
where dimensions dominated by behavioral patterns (high $ g_j $) retain more information from $ \mathbf{H}_b $, while those sensitive to temporal dynamics (low $ g_j $) emphasize $ \mathbf{H}_c $.  

To preserve original sequence information and enhance training stability, we apply residual connections and layer normalization:  
\begin{equation}
\mathbf{F}_{\text{fusion}} = \text{LayerNorm}(\mathbf{F}_{\text{fusion}} + \mathbf{H}_b + \mathbf{H}_c).
\end{equation}

Finally, we extend $ \mathbf{e}_x^{\text{session}} $ across the sequence length $ T $ and concatenate it with $ \mathbf{F}_{\text{fusion}} $ to obtain the final representation:  
\begin{equation}
\mathbf{F}_{\text{final}} = \text{Concat}(\mathbf{F}_{\text{fusion}},\ \mathbf{e}_\mathbf{x}^{\text{session}} \otimes \mathbf{1}_T),
\end{equation} 
where $ \mathbf{1}_T $ is an all-ones vector of length $ T $, and $ \otimes $ denotes the outer product.  

\subsection{ITD with Adaptive Threshold Optimization}
\subsubsection{MLP-based Probability Estimation}
The fused temporal-spatial representation $ \mathbf{F}_{\text{final}} \in \mathbb{R}^{T \times 2d_{\text{model}}} $ is passed through a Multi-Layer Perceptron (MLP) to estimate the anomaly probabilities. The probability sequence $ \mathbf{P} \in [0,1]^T $ is computed using two fully connected layers, with ReLU activation on the hidden layer and a sigmoid activation on the output layer. The process is defined as follows:
\begin{align}
\mathbf{H}_{\text{hidden}} = \text{ReLU}(\mathbf{F}_{\text{final}} \mathbf{W}_1 + \mathbf{b}_1), \\
\mathbf{P} = \sigma(\mathbf{H}_{\text{hidden}} \mathbf{W}_2 + \mathbf{b}_2),
\end{align} 
where $ \sigma(x) = \frac{1}{1 + \exp(-x)} $ is the Sigmoid activation function, which maps the input to a value between 0 and 1. Here, $ \mathbf{W}_1 \in \mathbb{R}^{2d_{\text{model}} \times H} $ and $ \mathbf{W}_2 \in \mathbb{R}^{H \times 1} $ are learnable weight matrices, where $H$ denotes the number of hidden units in the MLP. $ \mathbf{b}_1 $ and $ \mathbf{b}_2 $ are bias terms that help enhance the expressiveness of the model.

\subsubsection{Loss Function with Gating Regularization}
To optimize anomaly detection, we employ a hybrid loss function that integrates classification accuracy with feature fusion regularization. The objective function is formulated as:  
\begin{equation}
\mathcal{L} = \mathcal{L}_{\text{BCE}} + \lambda \mathcal{L}_{\text{G}}, \quad \lambda = 0.01,
\end{equation}  
where the Binary Cross-Entropy (BCE) loss ensures effective anomaly classification, and the gating regularization loss prevents the gating mechanism from collapsing into extreme values. The BCE loss is defined as:  
\begin{equation}
\mathcal{L}_{\text{BCE}} = - \frac{1}{T} \sum_{t=1}^{T} \left( y_t \log P_t + (1 - y_t) \log (1 - P_t) \right),
\end{equation}  
where \( y_t \) represents the ground truth label, and \( P_t \) is the predicted anomaly probability. This loss minimizes the divergence between predicted and actual labels, guiding the model to effectively distinguish between normal and anomalous behaviors.  
To ensure a balanced integration of behavioral and temporal interval features, we introduce the gating regularization loss, which is defined as:  
\begin{equation}
\mathcal{L}_{\text{G}} = \frac{1}{T d_{\text{model}}} \sum_{t=1}^{T} \sum_{j=1}^{d_{\text{model}}} G_{t,j} (1 - G_{t,j}).
\end{equation}  

This term discourages the gating values from collapsing into purely binary states (all 0s or all 1s), maintaining the model’s ability to adaptively weigh the importance of behavioral representations \( \mathbf{H}_b \) and temporal interval representations \( \mathbf{H}_c \). By setting \( \lambda = 0.01 \), the model preserves interpretability while ensuring that neither feature type dominates excessively.

\subsubsection{Adaptive Threshold Optimization}
\label{ATO}






To perform unsupervised anomaly classification at the user level, we adapt the classical Otsu’s thresholding strategy~(see Section~\ref{sec:otsu}) to the domain of per-user probability sequences. For each user $u$, we collect a daily sequence of anomaly scores $\mathbf{P} = [P_1, P_2, ..., P_T]$.

We first construct a normalized histogram $H(k)$ with 100 bins to approximate the empirical distribution of $\mathbf{P}$:

\begin{equation}
H(k) = \frac{1}{T} \sum_{t=1}^{T} \mathbb{I}\left(\left\lfloor \frac{P_t - \min(\mathbf{P})}{\max(\mathbf{P}) - \min(\mathbf{P})} \cdot 100 \right\rfloor = k\right),
\end{equation}

We then apply the Otsu criterion to this histogram to derive the optimal bin-level threshold $\tau^*$, and map it back to the probability space via:

\begin{equation}
\tau_u^* = \frac{\tau^*}{100} \cdot (\max(\mathbf{P}) - \min(\mathbf{P})) + \min(\mathbf{P}).
\end{equation}

This yields a dynamic, user-specific threshold that separates probable anomalies from benign activities. The final classification decision is given by:

\begin{equation}
y_t = 
\begin{cases}
1, & P_t \geq \tau_u^* \quad \text{(Anomaly)} \\
0, & P_t < \tau_u^* \quad \text{(Normal)}
\end{cases}
\quad \text{for } t \in \{1, 2, \dots, T\}.
\end{equation}
This user-aware binarization enables a data-driven thresholding mechanism that is sensitive to personalized behavioral patterns and avoids fixed or globally shared cutoffs.

\section{Experimental Setup}
\label{sec:setup}

\subsection{Datasets}
We utilize the CERT Insider Threat Test Dataset~\cite{lindauer2020insider}, an open-source dataset that provides comprehensive user activity logs and realistic insider threat scenarios. To evaluate the robustness and generalizability of our method, we employ both the r4.2 and r5.2 versions of the CERT dataset. Specifically, the r4.2 dataset contains 32,770,222 event records from 1,000 users, spanning from January 2010 to May 2011, with 7,323 labeled anomalous instances. The r5.2 dataset comprises 79,856,699 operations from 2,000 users, covering the period from January 2010 to June 2011, and includes 10,328 anomalies.
For behavior modeling, we organize user activities into daily session units and adopt an 8:2 split for training and testing. To address the inherent class imbalance in the dataset, we apply the Synthetic Minority Over-sampling Technique (SMOTE)~\cite{pal2023temporal} on the training set, thereby improving the representation of rare but critical anomalous behaviors.

\subsection{Comparison with Baseline Models}
To evaluate the Mamba framework, we compare it against a range of baselines, including deep learning models and state-of-the-art techniques. This includes time-series models like LSTM~\cite{villarreal2021hunting} and Transformer~\cite{vaswani2017attention}, which capture long-term dependencies; advanced methods such as Deep Isolated Forest (DIF)~\cite{cheng2023twostream} and Filter-Enhanced MLP (FMLP)~\cite{zhou2022filter}, which integrate traditional algorithms with deep learning; and leading models like DeepLog~\cite{du2017deeplog}, ITDBERT~\cite{huang2021itdbert}, OITP~\cite{manoharan2024optimising}, and CATE~\cite{xiao2024unveiling}, specializing in log data, insider threat detection, and anomaly detection in large-scale systems.

\subsection{Evaluation Metrics}
\label{Evaluation Metrics}
To assess the effectiveness of the proposed method, we utilize four key performance metrics: Precision, Recall, F1 Score (F1), and False Positive Rate (FPR)~\cite{pal2023temporal,xiao2024unveiling}. Precision indicates the accuracy of positive predictions, while Recall measures the sensitivity in detecting actual malicious activities. The F1 Score balances Precision and Recall for a comprehensive evaluation, and FPR assesses the rate of false alarms. Together, these metrics provide a thorough overview of the model's performance, highlighting its reliability and detection capabilities.

\subsection{Implementation}
\label{Implementation}
Our experiments were conducted on a platform running Windows 10, equipped with an Intel i7-11700 processor, 32GB of DDR4 RAM, and an NVIDIA GeForce RTX 3070 GPU with 8GB of VRAM. The machine learning-based models were implemented using the Python library Scikit-learn, while the deep learning-based models were developed using PyTorch. In our experiments, the Mamba network was configured with 2 layers, and the MLP detection layer was set to 3 layers. The parameter learning process can be found in \ref{Parameter Selection}.

\begin{table*}[!t]
    \footnotesize
    \centering
    \caption{Performance comparison of different algorithms.}
    \label{table: compare method}
    \begin{tabular}{lcccc|cccc}
    \toprule
    \multirow{2}{*}{\bfseries Method} & \multicolumn{4}{c}{\bfseries Cert r4.2} & \multicolumn{4}{c}{\bfseries Cert r5.2} \\
    \cmidrule(lr){2-5} \cmidrule(lr){6-9} 
    & Precision  & Recall  & F1  & FPR  & Precision  & Recall  & F1  & FPR  \\
    \midrule
    
    LSTM 
    & 77.32  & 74.32  & 75.79  & 12.62 
    & 79.87  & 78.41  & 79.13  & 16.46  \\  

    Transformer    
    & 79.32  & 75.22  & 77.22  & 19.40 
    & 82.87  & 80.24  & 81.53  & 19.69  \\  

    DIF   
    & 86.32  & 83.92  & 85.10  & 10.39 
    & 87.87  & 85.31  & 86.57  & 11.60  \\ 

    FMLP 
    & 83.32  & 81.32  & 82.31  & 18.74 
    & 86.87  & 84.71  & 85.78  & 12.19  \\  
    
    DeepLog 
    & 84.54 & 83.22 & 83.87 & 10.11  
    & 89.43 & 87.73 & 88.57 & 9.07  \\
    
    ITDBERT 
    & 80.77 & 75.32 & 77.95 & 18.99  
    & 83.89 & 80.41 & 82.11 & 13.70  \\
    
    OITP 
    & 77.25 & 70.69 & 73.82 & 13.46  
    & 82.41 & 80.38 & 81.38 & 12.12  \\
    
    CATE 
    & 91.32  & 85.71  & 88.43  & \textbf{4.85} 
    & \textbf{92.87}  & 89.41  & 91.11  & \textbf{4.92} \\  
    
    \midrule
    \bfseries \text{MambaITD} 
    & \textbf{91.79}   & \textbf{90.83}   & \textbf{91.31}    & 7.89 
    & 92.54   & \textbf{91.13}    & \textbf{91.83}    & 6.05 \\

    \bottomrule
    \end{tabular}
\end{table*}

\section{Experimental Results}
\label{sec:results}

\subsection{Comparison results}

Table~\ref{table: compare method} presents the performance comparison across multiple algorithms on Cert r4.2 and Cert r5.2. MambaITD achieves the highest F1-scores of 91.31 and 91.83, outperforming all baseline models in both datasets. This demonstrates its ability to effectively balance Precision and Recall while maintaining strong anomaly detection performance. Additionally, MambaITD exhibits a competitive False Positive Rate (FPR) of 7.89 and 6.05, outperforming models like Transformer (FPR = 19.40, 19.69) and FMLP (FPR = 18.74, 12.19), which suffer from higher false positives. While CATE achieves the lowest FPR (4.85 and 4.92), its F1-score is lower than MambaITD, indicating a potential trade-off where it reduces false positives at the cost of missing more true anomalies. DeepLog and ITDBERT maintain strong Recall but fail to achieve comparable F1-scores, highlighting MambaITD’s advantage in simultaneously achieving high Precision and Recall.

Compared to Transformer and DIF, which demonstrate good performance in Precision and Recall, MambaITD provides a more stable and robust detection capability, with a better balance across all metrics. The results indicate that MambaITD effectively reduces false positives while achieving superior detection accuracy, making it a more practical and efficient solution for insider threat detection.

\subsection{Ablation studies}
\begin{table}[!ht]
    \footnotesize
    \centering
    \caption{Performance comparison of ablation studies.}
    \label{table: ablation}
    \setlength{\tabcolsep}{5pt}
    \begin{tabular}{lcc|cc}
    \toprule
    \multirow{2}{*}{\bfseries Model} & \multicolumn{2}{c}{\bfseries Cert r4.2} & \multicolumn{2}{c}{\bfseries Cert r5.2} \\
    \cmidrule(lr){2-3} \cmidrule(lr){4-5} 
      & F1  & FPR   & F1  & FPR  \\
    \midrule
    
    MambaITD w/o ME + GFF + AT 
    & 81.34  & 13.57  & 83.42  & 16.21 \\  
    
    MambaITD w/o GFF + AT
    & 83.21  & 11.85  & 85.56  & 14.64 \\  
    
    MambaITD w/o ME + AT 
    & 84.12  & 10.72  & 86.32  & 13.81 \\  
    
    MambaITD w/o ME + GFF 
    & 82.58  & 14.01  & 84.67  & 15.35 \\  
    
    MambaITD w/o AT 
    & 85.02  & 9.14  & 87.76  & 12.59 \\  
    
    MambaITD w/o GFF 
    & 86.29  & 8.44  & 88.13  & 11.96 \\  
    
    MambaITD w/o ME 
    & 87.38  & \textbf{7.08}  & 89.54  & 10.75 \\  

    \midrule
    \bfseries MambaITD 
    & \textbf{91.28}  & 7.89  & \textbf{91.88}  & \textbf{6.05} \\  

    \bottomrule
    \end{tabular}
\end{table}

The ablation study results in Table~\ref{table: ablation} evaluate the impact of three key components—Mamba Encoder (ME), Gated Feature Fusion (GFF), and Adaptive Threshold (AT)—on the performance of MambaITD across the Cert r4.2 and Cert r5.2 datasets. The results highlight the contribution of each module in improving detection accuracy while reducing false positives.

\begin{itemize}
    \item \textbf{Baseline (w/o ME + GFF + AT):} Removing all components results in the lowest F1 scores (81.34, 83.42) and the highest FPR (13.57, 16.21), demonstrating the necessity of each module.
    
    \item \textbf{Single-Component Additions:}  
    \begin{itemize}
        \item Adding \textbf{ME} (w/o GFF + AT) improves F1 to 83.21 and 85.56, reducing FPR slightly, indicating that feature extraction alone enhances detection but lacks refinement.
        \item Adding \textbf{GFF} (w/o ME + AT) further boosts F1 to 84.12 and 86.32 while lowering FPR, highlighting its role in feature integration.
        \item Adding \textbf{AT} (w/o ME + GFF) provides limited improvement (F1: 82.58, 84.67), suggesting threshold optimization alone is insufficient.
    \end{itemize}

    \item \textbf{Two-Component Combinations:}  
    \begin{itemize}
        \item Keeping \textbf{ME and GFF} (w/o AT) improves F1 to 85.02 and 87.76, but FPR remains relatively high, showing that AT refines decision boundaries.
        \item Keeping \textbf{ME and AT} (w/o GFF) achieves F1 scores of 86.29 and 88.13, confirming GFF’s role in enhancing feature fusion.
        \item Keeping \textbf{GFF and AT} (w/o ME) achieves the best F1 among ablated models (87.38, 89.54) but still lags behind the full model.
    \end{itemize}

    \item \textbf{Full Model Performance:} The complete MambaITD model achieves the highest F1 scores (91.28, 91.88) and the lowest FPR (7.89, 6.05), confirming the effectiveness of integrating ME, GFF, and AT.
\end{itemize}

\subsection{Parameter Selection}
\label{Parameter Selection}

\begin{figure}[!t]
    \centering
    \subfigure[Mamba layer]{\includegraphics[width=0.44\linewidth]{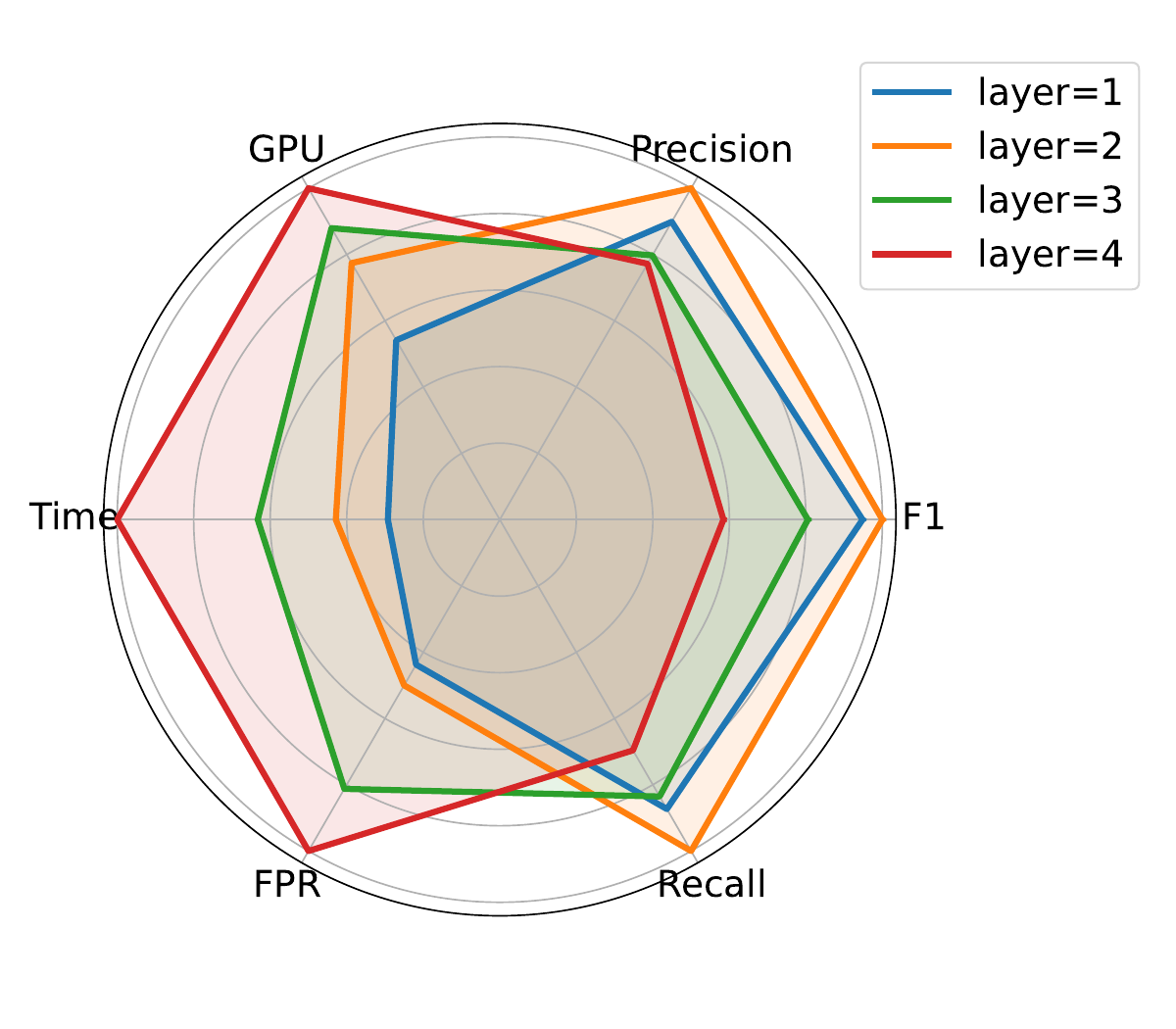}} 
    \subfigure[MLP layer]{\includegraphics[width=0.44\linewidth]{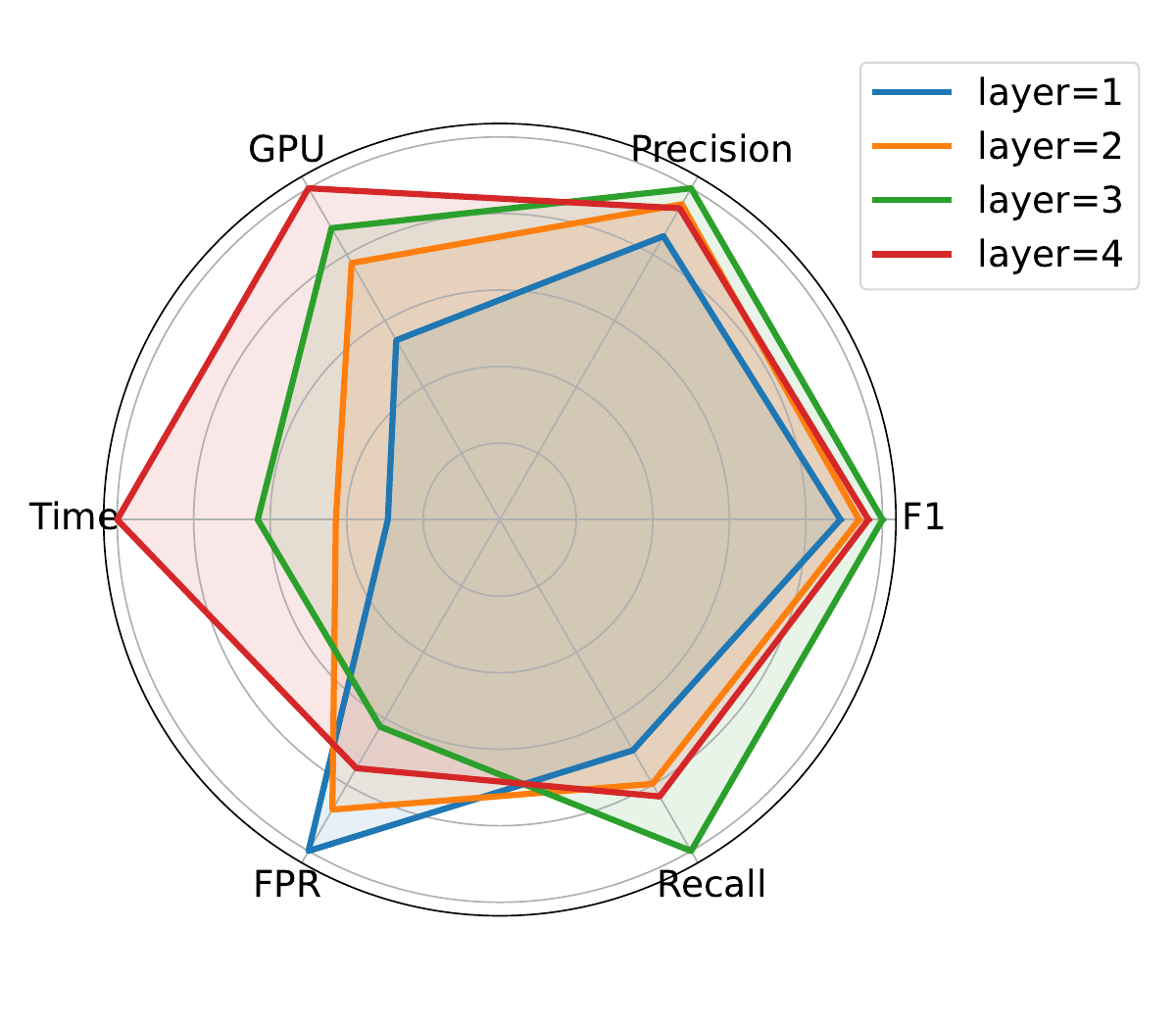}}

    \caption{Impact of Mamba and MLP Layer Variations on Model Performance.}
    \label{fig:parameter}
\end{figure}

As shown in the Fig.~\ref{fig:parameter}, we conducted experiments to explore the impact of different Mamba and MLP layer configurations on model performance. We evaluated the model using six key metrics. Notably, the left half of the radar chart (Time, GPU, and FPR) represents cost-related factors, where lower values indicate better efficiency. In contrast, the right half (Precision, F1, and Recall) reflects performance-based indicators, where higher values signify better results.

Observing the impact of Mamba layers, we find that when the number of layers is set to 2, the performance-oriented metrics (Precision, F1, and Recall) reach their optimal values, surpassing all other configurations. Additionally, the cost-related metrics (Time, GPU, and FPR) remain more favorable compared to layer settings of 1 and 4, making 2 layers the most balanced choice.

Similarly, for MLP layers, we observe that setting the MLP depth to 3 yields the best overall performance. Although the cost-related factors increase slightly, the significant improvement in Precision, F1, and Recall makes MLP with 3 layers the optimal configuration for achieving high accuracy while maintaining a reasonable trade-off with computational efficiency.

\subsection{Compatibility Analysis}
\begin{table}[!t]
    \footnotesize
    \centering
    \caption{Compatibility Analysis of different encoder model.}
    \label{table: architecture_comparison}
    \setlength{\tabcolsep}{4pt}
    \begin{tabular}{cccccc}
    \toprule
    \textbf{Dataset} & \textbf{Encoder Model} & \textbf{F1} & \textbf{FPR} & \textbf{Time(s)} & \textbf{GPU(MB)}\\
    \midrule
    \multirow{4}{*}{Cert r4.2} 
    & LSTM & 70.58 & 7.14 & 5.04 & 14.41\\
    & GRU & 77.17 & 7.14 & 4.78 & 11.16\\
    & Transformer & 87.50 & 2.38 & 3.71 & 9.46\\
    & Mamba & 87.50 & 2.38 & 2.73 & 8.20\\
    \midrule
    \multirow{4}{*}{Cert r5.2} 
    & LSTM & 76.63 & 6.31 & 7.51 & 14.83\\
    & GRU & 79.73 & 6.93 & 7.47 & 10.46\\
    & Transformer & 89.86 & 3.71 & 5.70 & 9.69\\
    & Mamba & 92.29 & 3.71 & 4.98 & 7.27\\
    \bottomrule
    \end{tabular}
\end{table}

To evaluate model compatibility, we conducted experiments on two user samples: ``AAF0535" from the Cert r4.2 dataset and BYO1846" from the Cert r5.2 dataset. Table~\ref{table: architecture_comparison} presents the results, comparing Mamba~\cite{gu2023mamba}, Transformer~\cite{vaswani2017attention}, LSTM~\cite{villarreal2021hunting}, and GRU~\cite{chung2014gated} in terms of F1 Score, False Positive Rate (FPR), computation time, and GPU usage.

The results show that Mamba and Transformer achieve the highest F1 Scores, with 87.50 on Cert r4.2 and 92.29 on Cert r5.2, outperforming LSTM (70.58, 76.63) and GRU (77.17, 79.73). In FPR, Mamba and Transformer maintain the lowest rates (2.38 and 3.71, respectively), significantly outperforming LSTM and GRU, which exhibit higher false positive rates (6.31–7.14).
Mamba surpasses Transformer in computational efficiency, running 1.36× faster (2.73s vs. 3.71s) on Cert r4.2 and 1.14× faster (4.98s vs. 5.70s) on Cert r5.2, translating to a 12.7\% reduction in runtime. In GPU memory consumption, Mamba also demonstrates better efficiency, requiring 8.20MB and 7.27MB, which is 13\% lower than Transformer (9.46MB and 9.69MB). In contrast, LSTM and GRU demand significantly more memory (10.46–14.83MB), further emphasizing their inefficiency.

Overall, Mamba matches Transformer in accuracy while significantly reducing computational cost and memory usage, making it a superior choice for large-scale, real-time anomaly detection in behavioral modeling.

\section{Conclusion}
\label{sec:conclusion}

In this study, we proposed a comprehensive approach to modeling multimodal user behavior logs to address the challenges posed by heterogeneous data sources and the complexity of sequential and non-sequential features. Our proposed framework integrates effective preprocessing, an innovative Mamba network encoder, and an adaptive threshold optimization mechanism, achieving significant improvements in both performance and interpretability. Results show that our approach achieves 1.14 times faster inference speed than Transformer, thereby reducing computation time by 12.7\%. These findings highlight the potential of our framework for applications in fraud detection and personalized recommendation systems, and the critical importance of efficient and interpretable analysis for building systems.

However, our study also has some limitations, including reliance on synthetic datasets, which may not fully capture the nuances of real-world behavior. Future work should focus on validating our approach using real user data and exploring further enhancements to the model, such as incorporating more advanced machine learning techniques or expanding the scope of user behavior analyzed.

\appendix

\subsection{Behavior Sequence Encoding Strategy}
\label{appendix:behavior_encoding} 

        
        
        
        

To ensure consistent representation of user behavior, we define a unique Mapping ID for each action by integrating three contextual attributes: Behavior (B), Device (D), and Time Segment (TS). As detailed in Table~\ref{tab:sequential_encoding_mixed_indexing}, our encoding strategy covers various log types with corresponding ID allocations across activity categories. Each action \( i \) is encoded using the following formula:

\begin{table}[!ht]
    \caption{Sequential Behavior Encoding Strategy.}
    \label{tab:sequential_encoding_mixed_indexing}
    \centering
    \footnotesize
    \setlength{\tabcolsep}{1pt}
    \begin{tabular}{l l l c}
        \toprule
        \textbf{Behavior (B)} & \textbf{Device (D)} & \textbf{Time Segment (TS)} & \textbf{ID Range} \\
        \midrule
        0--1: Logon / Logoff &
        \multirow{5}{*}{\shortstack[l]{0: Personal\\1: Department\\2: Supervisor\\3: Other}} &
        \multirow{5}{*}{\shortstack[l]{1: Working Hours\\2: Non-Working Hours}} &
        1--16 \\
        
        2--3: Connect / Disconnect & & & 17--32 \\
        
        4--15: File Operations      & & & 33--128 \\
        
        16--19: Email Operations    & & & 129--160 \\
        
        20--23: Web Activity        & & & 161--192 \\
        \bottomrule
    \end{tabular}
\end{table}

\begin{equation}
b_i = B_i \times (N_D \times N_{TS}) + D_i \times N_{TS} + TS_i, \quad i = 1, 2, \ldots, T,
\end{equation}

where \( B_i \) represents the behavior type, such as logon events, file operations, email exchanges, and web activities. The variable \( D_i \in \{0,1,2,3\} \) denotes the device category, distinguishing between personal, department-assigned, supervisor-assigned, and other devices. The variable \( TS_i \in \{1,2\} \) captures the temporal context, identifying whether the action occurred during working hours or non-working hours. This formulation guarantees that each behavior is mapped into a unique semantic-contextual index space, reflecting both operational intent and situational context.

To further refine the disambiguation of behaviors, we subdivide each behavior category into semantically meaningful subtypes. Specifically, for file operations (behavior codes 4–15), we account for two primary actions—open and write—applied across six common file formats: ZIP, DOC, PDF, EXE, TXT, and JPG, yielding a total of 12 distinct encoding positions. For email communications (behavior codes 16–19), we distinguish four types of message exchanges based on sender and recipient roles: Internal to Internal (I–I), Internal to External (I–E), External to Internal (E–I), and External to External (E–E). For web browsing activity (behavior codes 20–23), we classify accesses into four categories reflecting potential security concerns and benign behavior: cloud storage, hacktivist websites, job hunting, and neutral browsing.

By capturing this rich semantic and contextual structure, the proposed behavior encoding strategy offers a fine-grained yet compact representation of user activity sequences, thereby enabling effective learning of behavioral dynamics and facilitating accurate anomaly detection in downstream models.

\subsection{Statistical Feature Definitions}
\label{appendix:statistical_features}

\begin{table}[!ht]
    \caption{Statistical Features Derived from User Behavior, Device, and Time Segments.}
    \label{tab:statistical_features}
    \centering
    \setlength{\tabcolsep}{4pt}
    \footnotesize
    \begin{tabular}{llll}
        \toprule
        \textbf{Category} & \textbf{Factors} & \textbf{Info}  \\
        \midrule
        \multirow{5}{*}{\textbf{Behavior (B)}} 
        
        & Logon/Logoff & \multirow{12}{*}{\shortstack[l]{Count \\ \\Duration   } } \\
        & Connect/Disconnect &  \\
        & File Operations &   \\
        & Email Operations &   \\
        & Web Activity &   \\
        \addlinespace 
        \multirow{4}{*}{\textbf{Device (D)}}  
        & Personal &   \\
        & Department &   \\
        & Supervisor &  \\
        & Other &  \\
        \addlinespace 
        \multirow{2}{*}{\textbf{Time Segments (TS)}}  
        & Working Hours & \\
        & Non-Working Hours &   \\
        \bottomrule
    \end{tabular}
\end{table}

The statistical features are designed to capture user behavior patterns from three perspectives: Behavior (B), Device (D), and Time Segment (TS). Each feature is constructed by combining a specific factor from one of these categories with an associated statistical metric, forming a structured representation of user activities.

We adopt a \textit{Factor + Info} formulation, where each feature consists of a semantic aspect (e.g., Logon, File, Personal Device) and an associated statistic such as frequency (Count) or temporal span (Duration). For example, the feature \texttt{Logon + Count} quantifies the total number of logon events within a given session. This structured encoding facilitates fine-grained behavioral analysis across contextual dimensions.

Table~\ref{tab:statistical_features} summarizes the statistical features used in our framework.

\end{document}